# Application of dual-sensitive decision mechanism based on queuing theory in airport taxi management

*Fang He*[1,*]

*(School of Chemistry and Chemical Engineering, Xi'an University of Science and Technology, No.58 Yanta Middle Road, Wenyi Road Street, Beilin District, Xi 'an City)*

**Abstract**

With the rapid development of economy and global trade, flying has increasingly become the main way for people to travel, and taxi is the main transfer tool of the airport, especially in the airport with wide passenger sources and large passenger flow. However, at present, many large airports have the phenomenon of low taxi load rate, unbalanced income and long waiting time for passengers. Therefore, how can drivers make decisions to maximize their benefits, and how can management departments allocate resources and formulate management systems to improve ride efficiency and balance drivers' benefits have become urgent problems to be solved. This paper solves the above problems by establishing the taxi driver dual-sensitive decision model, the longitudinal taxi queuing model and the short-distance passenger re-return forced M/M/1 queuing model.

Problem 1 and Problem 2: Comprehensively considering the changing rule of passenger number and driver's income, using decision tree and decision matrix, the dual-sensitive decision model with cost and income co-exists is established. Taking Xiamen International Airport as an example, the rationality of the model and its dependence on related factors are verified by actual data.

Problem 3: In the case of two parallel lanes, the optimal layout is obtained by comparing and analyzing the three layout forms of vehicle driving path and pick-up point, and the longitudinal taxi queuing model is established to ensure the safety of vehicles and passengers and the maximum total ride efficiency by leaving the pick-up area on a curved cross path.

Problem 4: Taxi drivers can not choose passengers and refuse to take passengers, and drivers who carry short-distance passengers have less income. It is proposed to use the preoccupation mechanism to determine the priority of short-distance passengers returning to the taxi again, establish the preoccupation M/M/1 queuing model, and rationally arrange the priority to take passengers again, so that the benefits of drivers tend to be balanced.

simulink module is used to simulate and optimize the built model. Programming software is used to run the program code in the model and learn the historical data to verify the rationality and effectiveness of the built model, which makes the model more streamlined and further improves the robustness of the model. Finally, the model is evaluated and prospectively.

# A restatement of the problem

Most airports in China have separate channels for sending and receiving passengers, and taxi drivers who send passengers to the airport will face two choices: one is to go to the arrival area and queue up to take passengers back to the city. Taxis must go to the designated "car storage pool" to wait in line, according to the "first come, last come" to line up to pick up passengers, waiting time depends on the number of taxis and passengers in line, need to pay a certain time cost. The second is to return to the city directly empty for passengers. Taxi drivers pay empty fares and may lose potential revenue from carrying passengers.

According to the two choices drivers face, combined with the actual situation, a mathematical model is established to study the following problems:

(1) Comprehensively consider the changing law of airport passenger number and taxi driver's income, analyze the factors affecting taxi driver's decision making, establish the taxi driver's choice decision model, and give the corresponding choice strategy.

(2) By collecting relevant data of taxis in a domestic airport and the city where it is located, the selection scheme of taxi drivers in the airport is given, and the rationality of the built model and its dependence on relevant factors are analyzed.

(3) When there are two parallel lanes in the "boarding area" (taxi queuing for passengers and passengers queuing for rides), how should the management department set up the "boarding point" and arrange taxis and passengers reasonably, so as to maximize the overall boarding efficiency under the condition of ensuring the safety of vehicles and passengers?

(4) The passenger revenue of the airport taxi is related to the mileage of the passenger. The management department intends to give certain "priority" to some taxis that return to the airport for short-distance passengers, and design a feasible "priority" arrangement scheme to make the income of these taxis as balanced as possible.

## 2. Symbol and variable description

| symbol | explain |
| --- | --- |
| $\rho$ | The ratio of passenger arrival rate to service rate in the system |
| $\lambda$ | The average arrival rate of taxi passengers |
| C | The number of taxis waiting to pick up passengers in the storage pool |
| $\mu$ | The average service rate at a pick-up point |
| $C_\mu$ | The average service rate of the queuing system |
| Core(C,E); | The average service rate of the queuing system |

## 3. Model assumptions

1. The hypothesis does not consider the influence of the driver's personal special situation on the decision;
2. Assume that arriving flights are full of passengers;
3. Assume that the passenger's individual choice is not taken into account;
4. Assume that and are limited in the decision matrix;

## 4. Model building and solving

### 4.1 Problem 1

Taxi is a very large part of urban traffic [1], and the operation efficiency of taxi will greatly affect the operation of urban public transport. For this article, taxi drivers face two choices:

(1) Go to the arrival area and wait in line to pick up passengers and return to the city.

(2) Go directly to the city to collect passengers.

Based on the analysis of these two choices, the traditional decision tree and the cost-sensitive decision tree are used to introduce the cost-benefit dual-sensitive decision tree, and the driver's choice decision model is established to solve the driver's revenue- sensitive decision problem by comprehensively considering the changes in the number of passengers and the driver's income. The feasibility and effectiveness of the model is verified by an example [2].

### 4.1.1 A comparison of several decision tree algorithms

**(1) Traditional decision tree**

A decision tree is a tree-like structure that resembles a flow chart and can be used to make decisions. The construction of decision tree is completed by tree growth many times. Specifically, the training examples in a data set are constantly divided to form a tree structure. The decision tree is constructed by divide-and-conquer from the top down, and different styles of decision trees will be generated according to different split attribute selection strategies.

In traditional classification problem, given a training set $S = \{x_n, y_n\}_{n-1}^{N}$, $x_n$ belongs to the domain $X \subseteq R^D$, tag $y_n$ belongs to set, each sample $(x_n, y_n)$ independent from a collection of the unknown distribution $D = x \times y$, The task of the traditional classification algorithm is to use the training set S to discover a classifier g: x →y, and make a prediction Period error rate:

$$E(g) = \underset{(x,y) \sim D}{\varepsilon}[y \neq g(x)]$$

Minimum.

**(2) Cost-sensitive decision tree**

The traditional decision tree classification algorithm aims to maximize the accuracy of classification, or minimize the misclassification, which inevitably leads to the classification favoring the main class and ignoring the rare class or the class with little influence on the accuracy. In order to solve the decision problem in the cost-sensitive environment, the cost-sensitive decision tree is introduced based on the traditional decision tree [9].

The cost-sensitive algorithm assigns each example a cost vector $C \in [0, \infty]^k$, where *k* values are the costs incurred when example x is predicted to be *k*. A cost sensitive training set is given. $S_c = \{(x_n, y_n, c_m)\}_{n=1}^{N}$ that each cost sensitive training example $(x_n, y_n, c_n)$ is independently sampled on *a* set of unknown cost sensitive distributions $D_c = x \times y \times [0, \infty]^\lambda$. The cost as small as possible.

In the cost-sensitive decision tree algorithm[4], $C = (C_{ij})$ is commonly represented by the cost matrix, and (i, j) in $C = (C_{ij})$ indicates that the prediction is class i, which is actually the cost of class j. the total cost is:

$$L(x, i) = \sum_j P(j|x)C(i, j) \qquad (1)$$

If example x is predicted as class i and the above formula takes the minimum value, then the prediction at this time is called the best prediction.

**(3) Decision tree with Cost and Benefit (CBDT)**

Cost-sensitive decision tree takes minimizing cost as the goal to make decisions. Although this method can obtain decision-making results with low cost, it ignores the income problems brought. At the same time, low cost often fails to obtain high income. On the basis of the first two decision trees, this model comprehensively considers classification cost and classification benefit, and then establishes a decision tree model with the coexistence of cost and benefit, namely *CBDT* model [5].

Test cost: Set the test cost cost of node-splitting attribute in training data set *T* as standardized as :

$$TC(A_i)_{normal} = \frac{Max(1, TC(A_i))}{Max(1, TC(A_1), TC(A_2), TC(A_3) \cdots TC(A_n))} \quad (2)$$

Where the i th attribute in *attribute set*.

In the process of node category judgment, when a node is wrongly judged, it needs to pay a price, so the concept of cost matrix is defined.

Cost matrix: Let the nodes in the training data set T have n split attributes $a_1, a_2, a_3, \cdots, a_n$ with m distinct class identifiers $I_1, I_2, I_3, \cdots, I_m$, define its cost matrix C as: $C = (C_{ij})$, where represents the cost incurred when node k of class j is judged to be of class i. Im, is the correct classification when i=j, in this case $C_{ij}$=0, is the wrong classification when i≠j, in this case $C_{ij}$≠0.

Suppose that the category identification of node *k* in a dataset is {0,1}, and the cost of node is 10 when the node is wrongly judged, then the cost matrix of this node is:

$$C = \begin{Bmatrix} 0 & 10 \\ 10 & 0 \end{Bmatrix}$$

The cost of making a wrong decision is called a positive cost, and the cost of refusing a correct decision is called a false positive cost. Consider FP and FN together as misclassification costs. Conceptually, the cost of incorrectly marking an example should always be greater than the cost of correctly marking it. Yield matrix: Let the nodes in the training data set T have n split attributes $a_1, a_2, a_3 \cdots a_n$, with m distinct class identifiers $I_1, I_2, I_3 \cdots I_n$, given a node k, define its payoff matrix B as: $B = (b_{ij})$, where stands for the payoff generated when node k of class j is judged as class i. When i=j, for correct classification, $b_{ij}$≠0; For incorrect classification when i≠j, $b_{ij}$=0.

Put forward the concept of correctly classified revenue, and the revenue gained by making the correct decision is caled the revenue of the true sample. TR and DF are collectively referred to as positive categorical returns. Specific values for TR and DF are given by experts in the relevant field.

Conceptually, the gain from correctly labeling an example should always be greater than the gain from incorrectly labeling an example, and it is assumed that there is no gain from misclassifying an example. In order to balance benefits and costs, define the concept of benefits per unit cost.

Unit cost return: Let there be P positive examples and n negative examples in a node in the training data set T. defined as follows:

$$UCB(P) = \frac{P \times TR}{n \times FP}$$

When n* FP =0, take n* FP =1.

If the current node is judged to be a counterexample node(N), then the classification error cost is n * FP and the true classification benefit is n * DF, defined as follows:

$$UCB(P) = \frac{n \times DF}{p \times FN}$$

When P * FN =0, take P * FN =1.

Combine the above formulas to get: $UCB = \begin{cases} UCB(p) & \text{if } p \\ UCB(n) & \text{if } n \end{cases}$

**4.1.2 The establishment of decision tree model with coexisting cost and benefit**

Therefore, a decision tree with both cost and benefit is established. Because the goal of decision tree is to pursue the maximum income at the 6 minimum cost benefit. The steps are as follows:

1. Create a root node N;

2. If the training set is empty, return node N marked as Failure;

3. If every sample in the training set has the same class, return node N marked as that class;

4. If the candidate property is empty, return N as a leaf node marked with a class label;

5. for each candidate attribute list, select the split-attribute splitA in attribute _ list;

6. else recursively calls this algorithm.

**4.1.3 model solving**

In this paper, the model is used to make the most reasonable judgment in order to obtain greater benefits by considering the changing rules and benefits of airport passenger numbers. After comprehensive analysis of the topic and the model, the taxi driver's choice decision tree is abstracted, as shown in Figure 1.

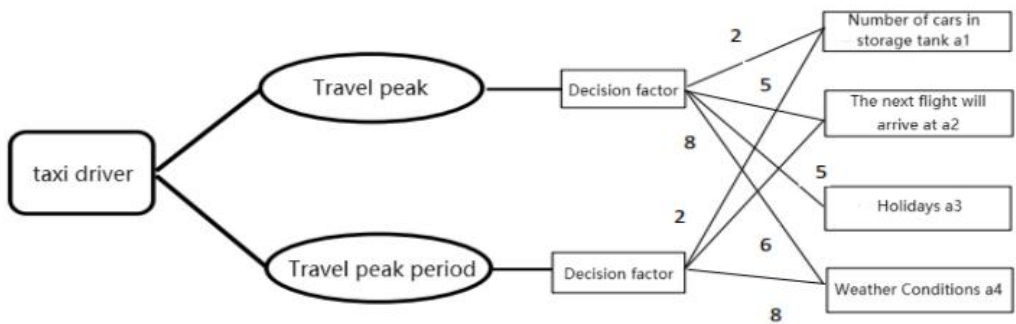

Figure 1. Driver choice decision tree

Let the taxi driver maximize the benefits to build a decision table as shown in the figure below:

Let $a_i(i = 1,2,3\cdots n)$ be the possible factor; $b_i(i = 1,2,3\cdots n)$ is the main factor $E_j$ (m< n) is the maximum expected value under the condition that $b_i$ is the main factor. Taking the maximum value in $E_j, \text{E} = \text{Max}\{E_j\}$, is the maximum return. The intersection of C and E is called a decision table. Let's call it Core(C, E);

Where:
$$a_i b_j = Random\{0,1,2,3\};$$
$$E_j = \sum_{i=0}^{n} a_i b_j; \quad \text{E} = \text{Max}\{E_j\};$$

The maximum expected value E can be calculated through the above decision matrix, which is conducive to the taxi driver to choose the best scheme and achieve the maximum return.

**4.2 Question Two**

With the improvement of people's living standard, airplane plays an increasingly important role in people's long-distance travel. Taxi is one of the most important means of transportation at the airport. Here, we comprehensively consider the factors affecting taxi revenue, such as flight schedule, travel peak period, holidays and so on. Therefore, how to adopt reasonable measures to enable taxi drivers to make correct decisions, so that they can obtain greater benefits has become the primary task of managers.

By collecting specific data and relevant taxi information of Xiamen (Yuanxiang) International Airport, this paper analyzes and studies the main factors affecting taxi drivers' income by using the model established in question1.

**1. Number and time of arrival of flights**

The key factor affecting taxi drivers' revenue is the number of passengers, which is largely determined by flight arrival time.

As you can see from the data we found, the airport has the largest number of flights and the corresponding number of passengers arriving during the three periods of 12:00- 13:00, 18- 19:00 and 23:00:24:00 in a day. The taxi mileage utilization rate is the highest during the first two periods, while the mileage utilization rate between 23:00 and 24:00 is relatively low. It can be seen that compared with weather conditions, travel peak periods and other factors, changes in passenger numbers play an important role in taxi drivers' decision making.

**2. The number of taxis**

The number of taxis is another major factor. For further research using the model, the following is the number and income of taxi companies and taxi drivers in Xiamen.

Table 1 Information table of airport flights and taxis

| Schedule | Flight arrivals | Passengers | The taxis is 10000 | Mileage ultization rate |
|---|---|---|---|---|
| 0 时-1 时 | 11 | 900 | 12.3 | 53% |
| 1 时-2 时 | 2 | 469 | 8 | 30% |
| 5 时-6 时 | 4 | 642 | 15 | 35% |
| 8 时-9 时 | 10 | 3523 | 23.78 | 44% |
| 9 时-10 时 | 12 | 5677 | 13.5 | 68% |
| 10 时-11 时 | 14 | 7834 | 24.63 | 92% |
| 11 时-12 时 | 15 | 8022 | 19 | 89% |
| 12 时-13 时 | 17 | 10229 | 17 | 98.90% |
| 13 时-14 时 | 11 | 968 | 24.98 | 75% |
| 14 时-15 时 | 14 | 6822 | 22 | 88.50% |
| 15 时-16 时 | 14 | 6655 | 13.5 | 97% |
| 16 时-17 时 | 16 | 8564 | 16.25 | 63% |
| 17 时-18 时 | 15 | 7812 | 24.69 | 89% |
| 18 时-19 时 | 18 | 15963 | 22 | 95% |
| 19 时-20 时 | 16 | 9541 | 19.44 | 77% |
| 20 时-21 时 | 11 | 789 | 16.25 | 69.80% |
| 21 时-22 时 | 10 | 564 | 17.69 | 57.89% |
| 22 时-23 时 | 12 | 1036 | 16.24 | 56.30% |
| 23 时-24 时 | 17 | 12387 | 12 | 44% |

As can be seen from Table 1, the airport has the largest number of flights in three periods of a day, from 12:00 to 13:00, from 18:00 to 19:00 and from 23:00 to 24:00. Since the flight information and the number of cars in the "car storage pool" are the information drivers know according to their experience, the waiting capacity of taxis in these three periods of a day is also the largest. It can be seen from this that compared with weather conditions, travel peak periods and other factors, the number of cars in the "storage pool" plays an important role in taxi drivers' decision-making.

**4.2.2 Model test**

By comparing the proportion of benefits under different decisions, the optimal solution is found and the maximum benefits are obtained. Its decision matrix and table are as follows:

|   | $a_1$ | $a_2$ | $a_3$ | $a_4$ |
|---|---|---|---|---|
| $b_1$ | $a_1b_1$ | $a_2b_1$ | $a_3b_1$ | $a_4b_1$ |
| $b_2$ | $a_1b_2$ | $a_2b_2$ | $a_3b_2$ | $a_4b_2$ |

$$C_E = \begin{bmatrix} a_1b_1 & a_2b_1 & a_3b_1 & a_4b_1 \\ a_1b_2 & a_2b_2 & a_3b_2 & a_4b_2 \end{bmatrix}$$

Obtained by $a_ib_j = Random\{0,1,2,3\}$: $a_ib_j = \{0,3,1,2; 1,3,0,1\}$;

According to $E_j = \sum_{i=0}^{n} a_ib_j$, $E = Max\{E_j\}$:

$$E_1 = 31; E_2 = 16; E = Max\{E_1, E_2\} = \{31,16\};$$

## 4.3 Question three

With the development of economic and international trade, airplane has increasingly become the main way for people to travel long distances. The traffic operation efficiency of airport has become the focus of people's attention. In particular, due to the unreasonable management mechanism of the airport management department or the specific environmental restrictions of the airport, the taxi passenger carrying efficiency is low and can not meet the needs of most passengers. Based on the analysis of taxi service efficiency and passenger queuing service in Xiamen International Airport, the paper discusses how to set passenger "pick-up point" and arrange taxis and passengers reasonably [6] when there are two parallel lanes, so as to achieve the highest overall ride efficiency and meet the ride demand of the vast majority of taxi passengers under the condition of ensuring safety [7].

### 4.3.1 A comparison of three kinds of taxi queuing service systems

1. Single-order taxi queuing service system

Single-point taxi queuing service system refers to the arrangement of a queue of passengers waiting for boarding corresponding to a boarding point [4]. Passengers and taxis line up in their respective waiting Spaces at the same time, one taxi corresponds to one customer, and only when a passenger leaves the queuing system after the service is finished can the passengers behind receive the service. The queuing model applicable to this system is M/M/1 queuing model. Its layout is shown in Figure 4.

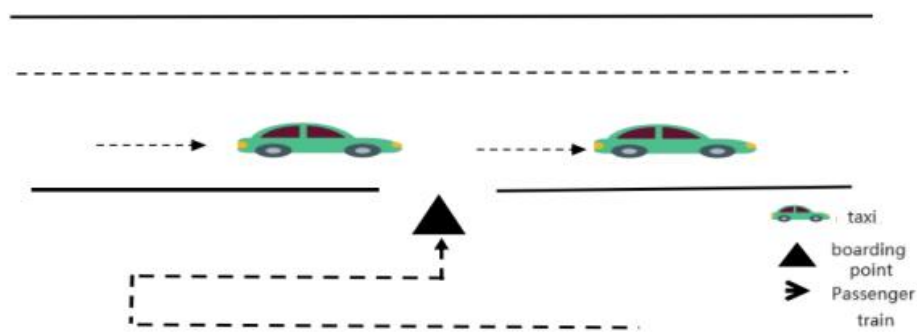

Figure 4 Single-point taxi queuing service system

During the peak travel period, the single-point taxi queuing service system is inefficient in carrying passengers, unable to meet the needs of a large number of passengers, and is not conducive to the improvement of airport traffic conditions. Therefore, the parallel lane taxi queuing system is proposed.

3. Parallel lane taxi queuing service system

The arrangement of each pick-up point in the parallel taxi queuing service system is in parallel. Passengers need to cross the taxi lane to each pick-up point after queuing. Although such a layout increases the pick-up points and improves the passenger efficiency to a certain extent, it is prone to passenger flow interference and passenger-vehicle conflict, which cannot guarantee the safety of vehicles and passengers, and prolongs the service time of the taxi pick-up points near the waiting line of passengers, thus affecting the passenger efficiency to a certain extent. This kind of taxi queuing system needs to set up a number of taxi lanes in the boarding area to increase the boarding point, and the system is suitable for M/G/1 queuing model. Its layout is shown in Figure 5.

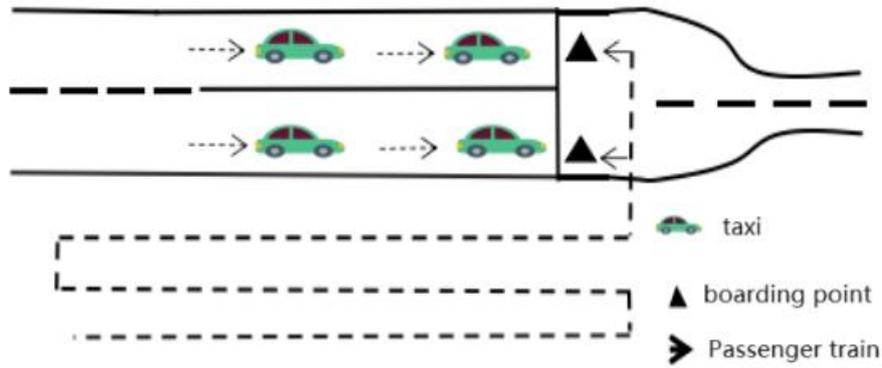

Figure 5 Parallel taxi queuing service system

Based on the disadvantages of parallel taxi queuing service system, which is easy to cause passenger flow interference and human-vehicle conflict, and can not guarantee the safety of vehicles and passengers, we put forward the longitudinal taxi queuing service system, which mainly solves the safety problems of vehicles and passengers.

**4. Longitudinal taxi queuing service system**

The longitudinal taxi queuing service system is a queuing system for multiple taxis and a public queue facing passengers. After passengers arrive at the queuing system according to certain arrival rules, the passengers at the front of the queue can be dispersed to the longitudinal "pick-up points" to receive services according to the current taxi service status of the pick-up point. Taxis will drive from the inside of the queue to the pick-up point for passenger service, and leave the pick-up point after the service is over, and then be replaced by the taxi behind them. Compared with the single-point taxi queuing service system, this kind of taxi queuing system increases the pick-up point and improves the passenger efficiency. After carrying passengers, the taxi drives across the route to another lane, which does not affect the vehicles that are carrying passengers, but also ensures the normal operation of its own service, and the most important thing is to ensure the safety of vehicles and passengers. Its layout form is shown in Figure 6.

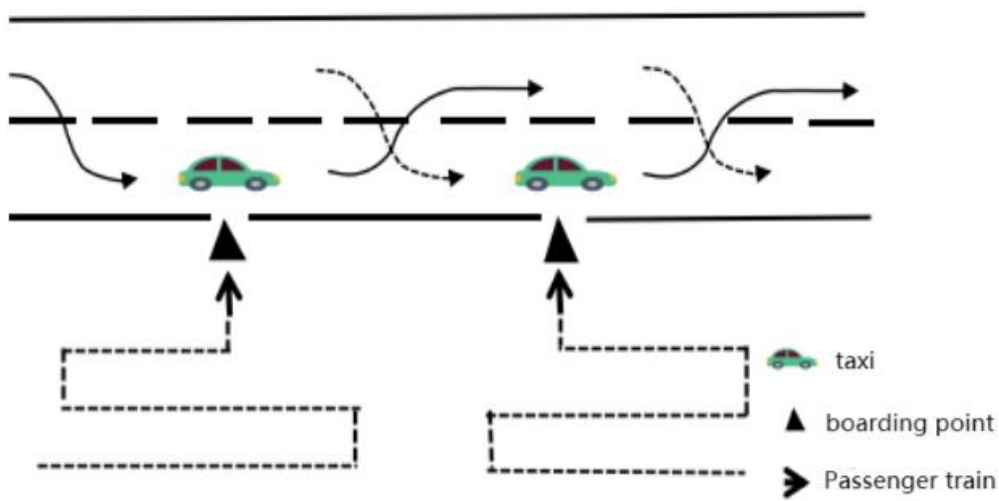

Figure 6 Multi-point parallel taxi queuing service system

Compared with the single-point taxi queuing service system, this kind of taxi queuing system increases the boarding point, improves the passenger efficiency, and the taxi driving after the passenger crossing route to the other lane drives forward, so that it does not affect the vehicle that is carrying the passenger, but also ensures that its service is normal, and the most important thing is to ensure the safety of the vehicle and passengers.

### 4.3.2 The longitudinal taxi queuing model

**1. queuing model**

According to the relevant theories of queuing theory [2], a queuing model is established for the queuing system in the passenger transport area of the airport. The proportion of passengers is:

$$P_0(C) = [\sum_{k=0}^{C-1} \frac{1}{k!}(\frac{\lambda}{\mu})^k + \frac{1}{C!}\frac{1}{1-\rho}(\frac{\lambda}{\mu})^c]^{-1} \qquad (3)$$

$$P_0(C) = \begin{cases} \frac{1}{n}(\frac{\lambda}{\mu})^k. & P_0(C)n = 1,2,\cdots,C \\ \frac{1}{C!C^{n-2}}(\frac{\lambda}{\mu}). & P_0(C)n = C+1 \end{cases} \qquad (4)$$

Using the captain $L_S$ of the passenger queue in the system and its stay time $W_S$ to analyze the system:

$$L_s = L_q + C_\rho = \frac{1}{C!}\frac{(C\rho)^C \rho}{(1-\rho)^2}P_0 + (\frac{\lambda}{\mu}) \qquad (5)$$

$$E(W_S) = \frac{P_n(C)}{C\mu(1-\rho)^2} = \frac{n\mu}{n!(n\mu-\lambda)^2}(\frac{\lambda}{\mu})^n P_0(C) \qquad (6)$$

**2. evaluation index of queuing system**

The evaluation indicators of a queuing system are queue length and passenger waiting time. The smaller the better, both from the perspective of the system and from the perspective of passengers. The queue length refers to the average number of passengers waiting in the queue in the system, and the captain refers to the sum of the queue length and the number of passengers receiving service. The waiting time refers to the average time from the passengers entering the "storage pool" to the beginning of receiving service, and the stay time refers to the average time from the passengers entering the "storage pool" to the passengers leaving the bus area.

The greater ρ is, the busier the queuing system is. When ρ is greater than 1, the system will be in the busy state of infinite queuing of customers. The busy rate ρ of the queuing system of the system can be calculated by the following formula:

$$\rho = \frac{\lambda}{C\mu} \qquad (7)$$

Where :C is the number of taxis, that is, the number of taxis waiting for passengers in the "car storage pool". λ is the average arrival rate of taxi passengers, that is, the number of passengers entering the queuing system per unit time; μ is the average service rate of a certain pick-up point, that is, the number of passengers who complete driving passengers out of the boarding area per unit time at a single service desk, and $C_\mu$ is the average service rate of the queuing system.

By comparing three kinds of taxi queuing service systems, the efficiency of the longitudinal queuing system in carrying passengers in two parallel lanes is highlighted. The queuing system sets up two parallel pick-up points on the same side of the lane, and carries passengers at the same time. Then it moves forward along the intersecting curve path, enters the free lane, and follows the straight line to the destination ground. Setting up pick-up points and arranging taxis and passengers in this way can not only improve the load rate, but also ensure the safety of vehicles and passengers [8].

**4.4 Question 4**

Queuing systems with priority [9] are generally divided into preoccupation systems and non-preoccupation systems. When the short-distance taxi with high priority arrives at the riding area, if it forces the process of the taxi with no priority to carry passengers, it is occupation; When a taxi with a high priority arrives at the ride- sharing area, if it does not force the ride-sharing process of a taxi with no priority, it is considered non- occupation. In this paper, the M/M/1 queuing system is used to study the priority of short-haul vehicles.

In the M/M/1 queuing system, the number of taxis arriving follows the Poisson distribution with parameter λ, and the average waiting length and average waiting time follow the exponential distribution with parameter μ. When non-short-distance taxis arrive at the ride-hailing area, they wait in the queue according to the rules of first come first served (FCFS); When short-haul taxis arrive at the ride area, wait in the short-haul fleet line according to the rules of first come first served (FCFS). t is the waiting time of any short-distance taxi in the queue, and its probability density function is f(t), then according to the queuing system theory [10], there is

$$f(t) = \begin{cases} P_0 & t = 0 \\ \sum_{n=1}^{\infty} P_0 \rho^n \Gamma_n, \mu(t) = \lambda P_0 e^{(\lambda-\mu)t} & t < 0 \end{cases} \quad (8)$$

Where:

$$\rho = \lambda/\mu; \; P_0 = 1 - \rho; \; \Gamma_{n,\mu}(t) = \frac{\mu^n t^{n-1} e^{-\mu t}}{(n-1)!} \quad (9)$$

Is the density function of the time required for n taxis to complete the passenger load. Suppose that the driver's yield is an exponential function of his waiting time.

$$S = e^{-\beta t} \quad (10)$$

Where s is the driver's benefit and β ≥0 is the sensitivity of driver's benefit to waiting time. As β ≥0, it can be found that the longer the driver waits, the lower the driver's earnings. According to the full probability formula, the expected value of the driver's earnings is as follows:

$$E(S) = \int_0^\infty e^{-\beta t}$$

$$= P_0 + \int_0^\infty e^{-\beta t} \lambda P_0 \, e^{(\lambda-\mu)t} dt = P_0 \frac{\mu+\beta}{\mu+\beta-\lambda} \quad (11)$$

For equation (11), then E=1. That is, when the driver is completely insensitive to the waiting time, its revenue will not be affected by the waiting time. In order to balance the benefits of different drivers, the management department uses the queuing system theory to realize the priority of short-distance taxis. Suppose that the time required by each passenger obees the same distribution, and let $ES = \frac{1}{\mu}, DS = \delta^2$ exist, that is, $ES = \int_0^\infty t dG(t) = \frac{1}{\mu}, DS = ES^2 - (\frac{1}{\mu})^2 = \delta^2$. Assuming that the service time series S required by various customers obey the same distribution G(t) = (t > 0). This paper takes the passenger carrying situation of airport taxi as the research theme. The arrival time and service time of taxi are very random, and it is impossible to know in advance what kind of distribution they will obey. The traditional analytical method is powerless to solve this situation, so this paper will adopt simulation method to study it.

The Monte Carlo simulation method is used to analyze the taxi passenger loading situation. First of all, random numbers that meet the specific distribution must be generated according to certain methods as the parameter input of the simulation model. Then the entity flow chart method is used to construct the system model to clarify the logical relationship between events, state changes and entity interactions, and the event priority scheduling method is used to construct the simulation model. Then run the simulation program on the computer, get the simulation results and statistical analysis and processing of the results. Finally, according to the simulation model output data analysis results, and make an evaluation of the "priority" arrangement plan.

1. **Build the simulation**

After the process analysis of the queuing model, the simulation model is built according to the event scheduling method. When using event scheduling method to build the simulation model, the event routine is the basic unit, and the system will continuously execute the corresponding event routine according to the sequence of events. The event routines here are divided into arriving taxi and leaving taxi. In order to be able to arrange different taxis reasonably, the model introduces a time control component and a logic control component. The time control unit is used to schedule the advance of the system clock, and the logic control unit is used to scan the schedule. By creating a schedule, record the taxis that will carry passengers in chronological order. The simulation flow chart is shown in Figure 7.

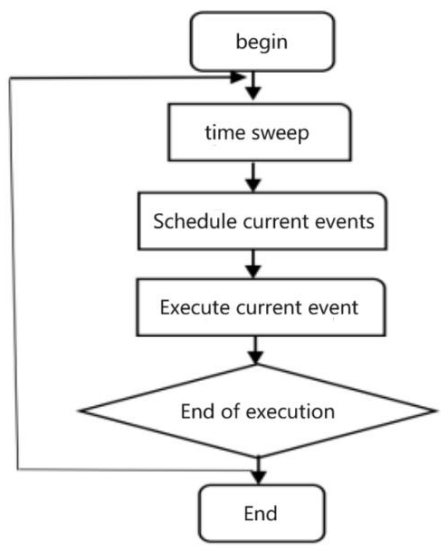

Figure 7 Simulation flow chart

In order to compare and analyze the two queuing models more scientifically, this paper adopts the relevant simulation output data analysis method. Considering that the initial conditions and parameters of the simulation run are the same, it can be assumed that the simulation results obey the normal distribution. According to classical statistical theory, for interval estimation of parameter X obeying normal distribution, it is necessary to distinguish the parameter population $\delta^2$; Known and unknown cases, the former using the standard normal distribution, the latter using the t distribution. Obviously, the variance of the population parameter to be estimated here is unknown, so we need to use the t distribution and take the sample variance as an approximation of the population variance to perform interval estimates of the parameters. The specific formula is as

follows:

$$\mu = \frac{1}{n}\sum_{j=1}^{n} X_j \pm t_{n.1.1\frac{a}{2}} \sqrt{S^2(n)/n} \qquad (12)$$

Where μ is the upper and lower limits of parameter interval estimation, $n$ is the sample data size (here is the number of data simulations), *and* is the sample variance. According to this formula, the 90% confidence intervals for the three variables 1, *W* and m and the degree of difference in the two scenarios can be calculated as follows:

$$0.0084 < 1 < 0.0309; 0.0091 < w < 0.0104; 1.2696 < m < 4.2304$$

Obviously, the confidence intervals for the variables 1 and m are on the large side. By increasing the simulation times and applying the program method to deal with it, the results are as follows:

$$0.0179 < 1 < 0.0218; 2 << 3$$

For the above two different priority claim queuing models, we first assume that the time interval of taxi arrival and the length of time required for passengers to get on the bus are uniformly distributed by (0, 1), and then establish simulation models respectively. The purpose of solving the queuing problem is to study the efficiency of the queuing service system, and there are two main indicators that reflect the quality of the system, namely, "captain" and "waiting time".

## 2. Simulation optimization

In order to further improve the fit of the simulation results, we generate random numbers through Poisson distribution and negative exponential distribution to simulate the permissible process of taxi forced queuing system in the ride-sharing area. The average waiting leader and average waiting time of all taxis in the simulation time were taken as indicators to measure the performance of the queuing system [11].

Suppose that the waiting leader of the system at time *t* is $L_q$, and the average waiting leader in the system is within the time [0,T], as follows:

$$L_q = \frac{1}{T}\int_0^T L\,dt \qquad (13)$$

Since integration cannot be realized in the actual simulation, we discretize $L_q$ as:

$$L_q = \sum_{t=1}^{n} \frac{X_i}{T} \qquad (14)$$

Where T is the system simulation time, n is the total number of customers arriving during the simulation time, and $t_1$ is when then $X_1$ is the product of the number of queues in the time interval $[t_{i-1}, t_i]$ and the length of the interval $[t_{i-1}, t_i]$.

Suppose that the average waiting time of customers in the system is $W_q$, and N is the time [0,T] to enter the system and receive the service $W_t$ is the waiting time of the i th taxi in the queuing system, then the taxi is in the queuing system[12].

The average waiting time is:

$$W_q = \frac{1}{N}\sum_{i=1}^{N} W_t \qquad (15)$$

From the above analysis, it can be seen that the number of taxis arriving at each time period in a day follows the Poisson distribution with parameter λ, and the average waiting length and average waiting time obey the negative exponential distribution with parameter u1 = 60.64 cars/hour and u2 = 28.46 cars/meter. According to the basic theory and formula of M/M/1 queue system, the average waiting probability, waiting leader and average waiting of taxi in different periods of the day were calculated by using the Stimulink module time, average length of stay. As shown in Table 2[13]:

Table 2 Information about short-distance taxis at different time periods in a day

| Time quantum | Delay probability | Average duration | Average waiting time(min) | Expected value of sojourn time (min) |
|---|---|---|---|---|
| 6:00-7:00 | 0.2695 | 0.3689 | 0.3819 | 1.4171 |
| 7:00-8:00 | 0.2719 | 0.3735 | 0.3866 | 1.4218 |
| 8:00-9:00 | 0.3285 | 0.4892 | 0.5064 | 1.5416 |
| 9:00-10:00 | 0.3509 | 0.5407 | 0.5597 | 1.5949 |
| 10:00-11:00 | 0.6729 | 2.0570 | 2.1294 | 3.1646 |
| 11:00-12:00 | 0.6743 | 2.0699 | 2.1428 | 3.1780 |
| 12:00-13:00 | 0.6132 | 1.5852 | 1.6410 | 2.6762 |
| 13:00-14:00 | 0.6632 | 1.9693 | 2.0386 | 3.0738 |
| 14:00-15:00 | 0.6774 | 2.0995 | 2.1734 | 3.2086 |
| 15:00-16:00 | 0.6901 | 2.2272 | 2.3056 | 3.3408 |
| 16:00-17:00 | 0.4082 | 0.6898 | 0.7141 | 1.7493 |
| 17:00-18:00 | 0.2723 | 0.3741 | 0.3873 | 1.4225 |
| 18:00-19:00 | 0.2046 | 0.2573 | 0.2663 | 1.3015 |

As can be seen from the above table, the waiting probability of short-distance taxi is higher during the period from 10:00 to 16:00, and the average captain, average waiting time and average stay time are also the longest. Therefore, short-distance taxi drivers have lower income than other drivers during this period. The number of taxis arriving in each period follows the Poisson distribution with λ parameter, and the average waiting time and average waiting length obey the negative exponential distribution with u1 = 60.64 cars/hour and u2 = 28.46 cars/meter. Considering the specific values of different times in the table, It can be concluded that the overall average income of short-distance taxi is u = 57.96, which is similar to the overall average income of other drivers. It can be seen that the forced M/M/1 queuing system used is suitable for short-distance taxi arrangements with priority, which can narrow the income gap and make the overall income of drivers tend to be balanced[14].

## 5. Model summary and evaluation

Aiming at the situation of few passengers and unbalanced profit of taxi in airport, the model is carried out to maximize the income of taxi drivers and improve the overall efficiency of passenger transport. The actual data is processed by decision matrix, and the fitting degree between the result and the model expectation is verified. In the case of two parallel lanes, the longitudinal taxi queuing model is established to rationally arrange the vehicle and passenger pick-up points in the driving area, and improve the overall passenger load rate while ensuring the safety of vehicles and passengers. In order to balance the income gap of taxis, the concept of preoccupation system is cited, and the relevant knowledge of queuing theory is combined to establish the preoccupation M/M/1 queuing model, and reasonable arrangement of taxi passengers is made, so that the revenue of taxis tends to be balanced.

At present, many large international airports in China generally have the problems of low taxi load rate and unbalanced income. The longitudinal taxi queuing model and the forced M/M/1 queuing model built in this paper can be used to solve these two problems respectively. The proposed model has great expansibility. If it is suitable for multiple parallel lanes, it is only necessary to establish a multi-point parallel taxi queuing system according to the existing model. At the same time, the dual-sensitive decision model with cost and benefit is suitable for many kinds of decisions, such as customer's decision in banking system, investment decision in financial industry and so on.